\documentclass[aps,pra,10pt,twocolumn,numerical]{revtex4-1}
\usepackage[utf8x]{inputenc}
\usepackage{amssymb,amsmath}
\usepackage{multirow}
\usepackage{graphics}
\usepackage{epsfig}
\usepackage{braket}
\usepackage{hyperref}
\usepackage{subfigure}

\begin{document}

\title{Ginzburg-Landau Theory for the Jaynes-Cummings-Hubbard Model}
\author{Christian Nietner$^1$ and Axel Pelster$^{2,3}$}

\affiliation{$^1$Institut f\"ur theoretische Physik, Technische Universit\"at Berlin, Hardenbergstr.36, 10623 Berlin, Germany\\
$^2$Hanse-Wissenschaftskolleg, Lehmkuhlenbusch 4, 27733 Delmenhorst, Germany\\
$^3$Fachbereich Physik und Forschungszentrum OPTIMAS, Technische Universit\"at Kaiserslautern, 67663 Kaiserslautern, Germany}
\begin{abstract}
\ We develop a Ginzburg-Landau theory for the Jaynes-Cummings-Hubbard model which effectively describes both static and dynamic properties of photons evolving in a cubic lattice of cavities, each filled with a two-level atom. To this end we calculate the effective action to first-order in the hopping parameter. Within a Landau description of a spatially and temporally constant order parameter we calculate the finite-temperature mean-field quantum phase boundary between a Mott insulating and a superfluid phase of polaritons. Furthermore, within the Ginzburg-Landau description of a spatio-temporal varying order parameter we determine the excitation spectra in both phases and, in particular, the sound velocity of light in the superfluid phase.
\end{abstract}

\maketitle

\section{Introduction}
In many fields of physics, especially in the fields of information processing, material science, superfluidity, and the relatively new field of quantum information theory, a profound understanding of strongly correlated quantum many-body systems is of striking importance in order to further improve existing applications and invent new ones \cite{Kay2004,Treutlein2006}. This is due to the fact that these research fields mainly use solid-state systems in which strongly correlated systems appear quite naturally. However, it is experimentally challenging to access the microscopic properties of such systems due to the short time and length scales involved. Therefore, motivated by Feynman's conjecture of the quantum simulator \cite{Feynman1982}, artificial structures have been considered to create effective many-body systems, which can be investigated under much more controllable and tunable experimental conditions.
\subsection{Optical Lattices}
The first attempts to build up artificial many-body structures used Josephson junction arrays \cite{VanDerZant1992,VanOudenaardenA1996}, which proved to be capable of simulating the dynamic properties of Bose-Hubbard systems including the generic quantum phase transition of this model \cite{Fisher1989}. Additionally, over the last two decades, the advances in preparing and controlling ultra-cold atoms led to new experimental realizations, which raised a huge amount of interest and research in this field. For example one investigated the interference of BEC clouds \cite{Naraschewski1996,Andrews1997}, studied rotating BEC's \cite{Madison2000,Abo-Shaeer2001,Kling2007,Kling2009}, observed spinor condensates \cite{Stamper-Kurn1998}, where BEC occurs in different hyperfine states, analyzed Bose-Fermi mixtures \cite{Schreck2001,Roethel2007}, where a pure BEC is contaminated with fermions, described BEC's in disordered potentials \cite{Pelster2006,Billy2008,Roati2008}, dipolar BEC's \cite{Gries2005,Aristeu}, and, more recently, tried to probe the properties of BEC's in zero gravity \cite{VanZoest2010}. \\
Of particular interest for the simulation of strongly correlated quantum many-body systems has been the realization of BEC's trapped in optical lattices \cite{Jaksch1998,Greiner2002}. Since their experimental realization, optical-lattice systems have initiated intensive studies and led to a multitude of new applications such as entanglement of atoms \cite{Jaksch1999,Mandel2003}, quantum teleportation \cite{Riebe2004}, Bell state experiments \cite{Roos2004}, disorder \cite{Sanchez-Palencia2005,Ahufinger2005,Clement2005,Krutitsky2006}, and ultra-cold molecules \cite{Yurovsky2000,Kokkelman2001}, to name but a few.
Unfortunately, the experimental approaches discussed so far face some crucial limitations. On the one hand, it is necessary to cool down the considered system to some nano Kelvin above absolute zero and, on the other hand, it is experimentally challenging to control and access single sites individually, However, recently developed experimental techniques also allow for single site addressability \cite{Greiner2010,Wuerz2009,Chin2009,Bloch2010}. Nevertheless, these experiments still need ultra-cold temperatures, a restriction which could be circumvented by using cavity lattices.
\subsection{Cavity QED Lattices}
Encouraged by the latest progress in the fabrication and manipulation of micro cavities \cite{Vahala2003,Khitrova2006,Benson2006}, Philippe Grangier and others \cite{Grangier2000,Matter2000,Lukin,Hartmann2006,Greentree2006,Angelakis2007} proposed a new experimental setup using cavity quantum electrodynamics (QED) schemes.
The underlying idea behind this new approach is to build up a lattice from micro cavities and place some real or artificial atoms in each cavity, for example Josephson junctions or quantum dots. Subsequently, light is coupled into the system in such a way, that it interacts with the atoms. As a result, the coupling between the light field and the atoms leads to the formation of quasi-particles, so called polaritons. These quasi-particles behave like real bosonic particles on the lattice. In fact, Bose-Einstein condensation of polaritons was recently experimentally achieved in semi-conductor cavities filled with quantum wells \cite{Deng2002,Kasprzak2006, Balili2009} and even superfluidity could be observed \cite{Amo2007,Amo2009}.
This new idea for a quantum simulator based on cavity QED does not share the above mentioned limitations of the optical lattice approach. Due to relatively huge distances between the cavities, local control and accessibility emerges quite naturally for these systems. Hence, it is possible to analyze these systems without destroying them, in contrast to the time-of-flight imaging technique used for ultracold quantum gases. Since the atoms are trapped inside the cavities right from the start and their thermal motion does not quantitatively disturb the polariton dynamics \cite{Muradyan2001}, BEC experiments with cavity QED setups can thus be performed even at room temperatures \cite{Sun2010}. However, in order to facilitate stable experiments with this setup, one needs a strong coupling between light and matter in order to reduce the losses induced by the spontaneous emission. Fortunately, over the past few years, this so called \emph{strong-coupling regime} \cite{Laussy2010,Laucht2010,Carmon2005} has become experimentally accessible for a large number of different setups \cite{Sun2010,Ctistis2010,Spillane2003,Brossard2010,Wallraff2004,Kippenberg2004,Bajcsy2009,Trupke2007}.\\
However, a drawback of the strong coupling is the short polariton lifetime which prevents to reach thermal equilibrium. Thus, in a real experiment the polaritons decay faster then they can equilibrate via phonon emission and polariton-polariton scattering and therefore the system needs to be continuously pumped by an external laser. Recently, F. Nissen et al.~theoretically investigated the coherently driven and dissipative JCHM \cite{Schmidt2012}. They could show that the important photon-blockade effect prevails in this system for weak hopping. This effect guarantees a strong polariton-polariton interaction which allows the system to reach a quasi thermal equilibrium \cite{Deng2002,Kasprzak2006}. In principal one can modify the method presented in this paper to include external baths in order to account for this non-equilibrium situation. Subsequently tracing out the bath degrees of freedom yields an effective theory for the reduced system. Nevertheless, we will in this paper focus on the quasi equilibrium situation with a steady state polariton density in order to demonstrate the theoretical feasibility.
\subsection{Outline of the Paper}
This paper is structured as follows. In Section $2$ we will introduce the underlying Hamiltonian for the Jaynes-Cummings Hubbard (JCH) model and briefly analyze its dynamical properties in some special limits. In Section $3$ we will derive an effective Ginzburg-Landau action for the model. To this end we will use the approach of Refs.~\cite{key-1,Santos2009}, that has already been successfully applied to analyze collective excitations of Bose gases in optical lattices \cite{Grass2011,Grass2011a}, and transfer it to a cavity QED system. In contrast to other methods, this procedure yields a consistent thermodynamic theory for finite temperatures in the whole quantum phase diagram. Therefore our Ginzburg-Landau approach is of relevance in order to develop a thermometer for the JCHM. For instance, to establish a thermometer for the usual Bose-Hubbard model turned out to be a difficult task \cite{Hoffmann,McKay}. The respective results for both static and dynamic properties as well as their temperature dependence obtained from the effective action will be presented in Section $4$. In particular we focus on the excitation spectra and on the sound velocity of light and discuss their dependence on the experimentally accessible detuning parameter.
\section{The model}
For our lattice model we consider a Hamiltonian of the form
\begin{equation}\label{E:GeneralJCHHamiltonian}
 \hat{H}=\hat{H}_{0}+\hat{H}_{\rm{h}}.
\end{equation}
Using the convention $\hbar=1$ throughout this paper, the local part of this Hamiltonian is given by
\begin{equation}\label{E:JaynesCummingsHamiltonian}
 \hat{H}_{0}=\underset{i}{\sum}\left[ (\omega-\mu)\,\hat{n}_{i}+\Delta\,\hat{\sigma}^{+}_{i}\hat{\sigma}^{-}_{i}+ g\left(\hat{a}_{i}\,\hat{\sigma}^{+}_{i}+\hat{a}^{\dagger}_{i}\,\hat{\sigma}^{-}_{i}\right)\right]
\end{equation}
and consists of a sum over local Jaynes-Cummings Hamiltonians in the \textit{rotating wave approximation} (RWA) \cite{RWA}. Here $\omega$ denotes the frequency of the local monochromatic photon fields associated with the annihilation and creation operators $\hat{a}_{i}$ and $\hat{a}^{\dagger}_{i}$. The parameter $\Delta=\omega-\epsilon$ corresponds to the detuning between the local mode frequency $\omega$ and the energy splitting $\epsilon$ of each two-level system. The operators $\hat{\sigma}^{+}_{i}$ and $\hat{\sigma}^{-}_{i}$ are the ladder operators of the two-level systems and $g$ is the real coupling strength between each two-level system and the bosonic mode. The operator $\hat{n}_{i}=\hat{a}^{\dagger}_{i}\hat{a}_{i}+\hat{\sigma}^{+}_{i}\hat{\sigma}^{-}_{i}$ represents the on-site polariton number operator and $\hat{N}=\sum_{i}\hat{n}_{i}$ is the total polariton number operator. As it commutes with $\hat{H}_{0}$, the number of these polaritons, which are coupled excitations of the two-level system and the local light fields, is a conserved quantity in this model. Finally, working in the grand-canonical ensemble, yields an additional term in the Hamiltonian which is proportional to the polariton number operator and the chemical potential $\mu$.\\
The Jaynes-Cummings model is well known in the literature and the underlying Hamiltonian can be exactly diagonalized within the RWA \cite{Jaynes} leading to the energy eigenvalues
\begin{eqnarray}
\label{E:JC_EigenValues}
E_{n\pm}&=&-\mu_{\rm{eff}}\,n+\frac{1}{2}\left[\Delta\pm R_{n}(\Delta)\right],\;n>1, \\E_{0}&=&0,\; n=0.
\end{eqnarray}
\\
Here, we introduced the generalized Rabi frequency $R_{n}(\Delta)=\sqrt{\Delta^{2}+4\,g^{2}n}$ and the effective chemical potential $\mu_{\rm{eff}}=\mu-\omega$. The eigenvalues $E_{n\pm}$ correspond to the energy eigenstates
\begin{subequations}\label{E:DressedStatesDefinition}
\begin{align}
 \Ket{n,+}&=\sin\theta_{n}\ket{n,g}+\cos\theta_{n}\ket{n-1,e}\,,\\
 \Ket{n,-}&=\cos\theta_{n}\ket{n,g}-\sin\theta_{n}\ket{n-1,e}\,,
\end{align}
\end{subequations}
with the mixing angle $\theta_{n}=\frac{1}{2}\arctan\left(\frac{2\,g\sqrt{n}}{\Delta}\right)$. The vacuum state is given by $\ket{0}=\ket{0,g}$. We note that the energy spectrum for a fixed polariton number $n>0$ naturally splits into an upper and a lower branch, where the lower branch is always lower in energy than the upper branch.\\
The possibility for photons to tunnel between next neighboring cavities is modeled by a Hubbard-like hopping Hamiltonian of the form
\begin{equation}\label{E:HoppingHamiltonian}
 \hat{H}_{\rm{h}}=-\underset{<i,j>}{\sum}\kappa_{i,j}\,\hat{a}^{\dagger}_{i}\hat{a}_{j},
\end{equation}
where the sum runs over all next neighbor lattice sites. As the tunnel matrix elements $\kappa_{i,j}$ exponentially decay with increasing distance between the lattice sites $i$ and $j$, it is justified to assume that $\kappa_{i,j}=\kappa$ if $i$ and $j$ are next neighbors and  $\kappa_{i,j}=0$ otherwise.\\
As a first analysis of the ground-state of the model described by Eq.~\eqref{E:GeneralJCHHamiltonian} we consider the physically relevant extremes of both the atomic and the hopping limit. In the atomic limit $\kappa\ll g$, Eq.~\eqref{E:GeneralJCHHamiltonian} simplifies to $\hat{H}\thickapprox\hat{H}_{0}$ which decomposes into purely local contributions with eigenvalues \eqref{E:JC_EigenValues}. Obviously, in this regime the photons can not move in the lattice and, thus, all excitations are pinned to their respective lattice sites. Therefore, the ground state wave function of the whole lattice is simply a direct product of the local on-site ground state wave functions. For this reason the ground state of the whole system is reached when each on-site Jaynes-Cummings system is in the lowest energy state $E_0$ or one of the lower-branch states $E_{n-}$. However, decreasing the difference $\omega-\mu$ one eventually reaches a point when $E_{0}=E_{1-}$ and, hence, adding a polariton excitation becomes energetically favorable. The successive repetition of this argument leads to a complete set of such degeneracy points $E_{n-}=E_{(n+1)-}$, which are characterized by the explicit relations:
\begin{subequations}
\begin{align}
 \frac{\mu_{\rm{eff}}}{g}&=\frac{1}{2\,g}\left[R_{n}(\Delta)-R_{n+1}(\Delta)\right],\;&n>1\\
 \frac{\mu_{\rm{eff}}}{g}&=\frac{1}{2\,g}\left[\Delta-R_{1}(\Delta)\right],\;&n=0.
\end{align}
\end{subequations}
From the above discussion follows that, in the atomic limit, the local polariton number is fixed at each lattice site for a given set of parameters. When the total polariton number equals an integer multiple of the number of cavities, this regime is called the Mott insulating phase which has been predicted \cite{Fisher1989} and experimentally observed in the Bose-Hubbard model \cite{Bloch2009}. \\
As a second interesting limit we investigate the ground-state in the regime when $\kappa\gg g$, i.e. where the photon hopping dominates the system dynamics.
We additionally assume that all two-level systems are in their respective ground-state as the system minimizes the energy. Hence, we can drop all atomic contributions, which leads to the hopping-limit Hamiltonian
\begin{equation}
   \hat{H}\thickapprox-\mu_{\rm{eff}}\hat{N}-\underset{<i,j>}{\sum}\kappa_{i,j}\,\hat{a}^{\dagger}_{i}\hat{a}_{j}.
\end{equation}
This Hamiltonian can be diagonalized in Fourier space leading to
\begin{equation}\label{E:ReducedHoppinglimitHamiltonian}
 \hat{H}=\underset{\textbf{k}}{\sum}\epsilon(\textbf{k})\,\hat{a}^\dagger_{\textbf{k}}\hat{a}_{\textbf{k}},
\end{equation}
with the energy dispersion $\epsilon(\textbf{k})=-\mu_{\rm{eff}}-2\,\kappa\,\overset{d}{\underset{i=1}{\sum}}\cos(k_{i}\,a)$, where $a$ is the lattice constant of a simple $d$-dimensional cubic lattice. Thus, the Hamiltonian of the hopping limit \eqref{E:ReducedHoppinglimitHamiltonian} is local in Fourier space. This situation corresponds to the superfluid phase of the system. Signatures of this phase have already been observed in the interference patterns of time-of-flight experiments with Bose-Hubbard systems \cite{Bloch2008}.
\section{Effective Action}
In the following section we derive an effective action for the JCH model from the free energy. Explicitly calculating the lowest hopping order contributions of the effective action amounts effectively to a resummation of infinitely many hopping contributions. Therefore, this effective action allows us to determine the quantum phase transition as well as calculate the excitation spectra, energy gap, effective mass, and sound velocity for finite temperatures in the Mott phase and in the superfluid phase.
\subsection{Free Energy}
Following the approach used for example in Refs.~\cite{key-1,Santos2009} we additionally introduce source currents $j(\tau),\,j^{*}(\tau)$, in the system Hamiltonian \eqref{E:GeneralJCHHamiltonian} leading to the new Hamiltonian
\begin{equation}\label{E:CurrentHamiltonian}
 \hat{H}^{\prime}(\tau)\left[j,j^{*}\right]=\hat{H}+\sum_{i}\left[j_{i}^{*}(\tau)\,\hat{a}_{i}+j_{i}(\tau)\,\hat{a}_{i}^{\dagger}\right],
\end{equation}
which is now a functional of the currents. These artificial currents, which explicitly depend on the imaginary-time variable $\tau$, will be used to artificially break the $U(1)$ symmetry that is responsible for the quantum phase transition in the model \cite{Sachdev}. Since all physical results are obtained in the limit of vanishing currents, we can treat all terms in the Hamiltonian \eqref{E:CurrentHamiltonian}, which are proportional to the source currents, as small quantities. Furthermore, for small hopping amplitudes $\kappa_{ij}$ all off-diagonal contributions in Eq.~\eqref{E:CurrentHamiltonian} become small. Hence, we decompose the new system Hamiltonian into the form
\begin{equation}
 \hat{H}^{\prime}(\tau)\left[j,j^{*}\right]=\hat{H}_{0}+\hat{H}_{1}(\tau)\left[j,j^{*}\right],
\end{equation}
where the local part $\hat{H}_{0}$ from Eq.~\eqref{E:JaynesCummingsHamiltonian} is exactly solvable and the remaining part
\begin{equation}\label{E:PertubationHamiltonian}
 \hat{H}_{1}(\tau)\left[j,j^{*}\right]=\hat{H}_{\rm{h}}+\sum_{i}\left[j_{i}^{*}(\tau)\,\hat{a}_{i}+j_{i}(\tau)\,\hat{a}_{i}^{\dagger}\right]
\end{equation}
with the hopping Hamiltonian $\hat{H}_{\rm{h}}$ from Eq.~\eqref{E:HoppingHamiltonian} can be treated as a perturbation.\\
We aim at establishing a thermodynamic perturbation theory in the present section. Therefore, it is convenient to switch from the Schr\"odinger picture to the imaginary-time Dirac interaction picture. This leads to a reformulation of the partition function as a perturbation series, involving just the perturbative part \eqref{E:PertubationHamiltonian} of the full Hamiltonian of the system. Introducing the abbreviation $\left\langle \bullet\right\rangle _{0}=\frac{1}{\mathcal{Z}_{0}}\textrm{Tr}\left\{ \bullet\; e^{-\beta\,\hat{H}_{0}}\right\}$ with the inverse temperature $\beta=1/(k_{\rm{B}} T)$ for the thermal average with respect to the unperturbed system, the partition function takes on the following form
\begin{equation}\label{E:ZasThermalAverage}
 \mathcal{Z}=\mathcal{Z}_{0}\left\langle \hat{U}_{\rm D}(\beta,0)\right\rangle_{0}.
\end{equation}
Here, the partition function of the unperturbed system is given by
\begin{equation}\label{E:DefinitionZzer0}
 \mathcal{Z}_{0}=\textrm{Tr}\left\{ e^{-\beta\,\hat{H}_{0}}\right\}
\end{equation}
and the imaginary-time evolution operator in the Dirac picture is defined as
\begin{equation}\label{E:DefinitionImaginaryTimeEvolutionOperator}
 \hat{U}_{\rm D}(\beta,0)=\hat{T}\exp\left\lbrace -\int_{0}^{\beta}d\tau\,\hat{H}_{1}(\tau)\left[j,j^{*}\right]\right\rbrace.
\end{equation}
Note that all operators, which depend on imaginary-time variables, have to be taken in the imaginary-time Dirac interaction picture, i.e. $\hat{O}_{\rm D}(\tau)=e^{\hat{H}_{0}\tau}\hat{O}e^{-\hat{H}_{0}\,\tau}$.\\
Using the above definition \eqref{E:PertubationHamiltonian} together with Eqs.~\eqref{E:ZasThermalAverage} and \eqref{E:DefinitionImaginaryTimeEvolutionOperator} we see that the partition function also becomes a functional of $j(\tau)$ and $j^{*}(\tau)$. Splitting the grand-canonical partition functional into the respective perturbative contributions
\begin{equation}
 \mathcal{Z}\left[j,j^{*}\right]=\mathcal{Z}_{0}\left\lbrace 1+\sum_{n=1}^{\infty}\mathcal{Z}_{n}\left[j,j^{*}\right]\right\rbrace ,
\end{equation}
the free energy defined by $\mathcal{F}=-\beta^{-1}\ln\mathcal{Z}$ can be written as
\begin{equation}\label{E:FreeEnergyApprox1}
 \mathcal{F}\left[j,j^{*}\right]=\mathcal{F}_{0}-\frac{1}{\beta}\ln\left\{1+\sum_{n=1}^{\infty}\mathcal{Z}_{n}\left[j,j^{*}\right]\right\}\,.
\end{equation}
Here we introduced the free energy of the unperturbed system as the usual expression $\mathcal{F}_{0}=-\beta^{-1}\ln\mathcal{Z}_{0}$. In the next step we expand the free energy functional in a power series of the perturbation parameters $j,j^*,\kappa_{ij}$. In this paper we will focus just on the lowest order contributions from the hopping and thus neglect all terms of higher than first order in $\kappa_{ij}$. Furthermore, according to the Landau theory \cite{Landau}, one needs to consider all terms at least up to fourth order in the order parameter to describe the thermodynamic properties of a second order phase transition. Since we will see later on that the source currents $j,j^*$ are of the order of the Ginzburg-Landau order parameter for the considered system, we thus have to calculate the power series up to fourth order in $j,j^*$. Hence, we expand the logarithm in expression \eqref{E:FreeEnergyApprox1} and keep all terms up to fourth order in $j$ and $j^*$ and first order in $\kappa$.\\
This procedure leads to an expansion of the free energy functional in terms of imaginary time integrals over sums of products of thermal Green functions with respect to the unperturbed system of increasing order. The $n$th order thermal Green function with respect to the unperturbed system is defined as
\begin{align}\label{E:DefinitionOfNParticleGreensFunction}
 G_{n}^{(0)}&\left(\tau_{1}^{\prime},i_{1}^{\prime};\ldots;\tau_{n}^{\prime},i_{n}^{\prime}|\tau_{1},i_{1};\ldots;\tau_{n},i_{n}\right)\notag\\
 &=\left\langle \hat{T}\left[\hat{a}_{i_{1}^{\prime}}^{\dagger}(\tau_{1}^{\prime})\,\hat{a}_{i_{1}}(\tau_{1})\ldots\hat{a}_{i_{n}^{\prime}}^{\dagger}(\tau_{n}^{\prime})\,\hat{a}_{i_{n}}(\tau_{n})\right]\right\rangle _{0}\,.
\end{align}
In principle, one could now make use of the definition \eqref{E:DefinitionOfNParticleGreensFunction} and calculate the expansion coefficients of the free energy straightforwardly. However, with increasing order of the thermal Green function the calculation becomes more and more complex due to the increasing number of space- and time-index permutations. Therefore, we use another approach to calculate the thermal Green functions, which automatically takes care of the emerging problems.

\subsection{Cumulant Expansion}

Usually one would apply in field theory the Wick theorem to decompose $n$-point correlation functions into sums of products of $2$-point correlation functions \cite{Negele}. Unfortunately, this is not possible for the considered system, since the Wick theorem just holds for systems, where the unperturbed Hamiltonian is linear in the occupation number operator. Instead, in our case one has to use the so called \textit{cumulant expansion}, which was originally developed for the Hubbard model as is reviewed by Metzner \cite{Metzner1991}. It states that the logarithm of the partition function is given by the sum of all connected Green functions. The power of this approach lies in the fact that these connected Green functions can subsequently be derived from a single generating functional by performing functional derivatives with respect to the currents.  Due to the fact that the unperturbed Hamiltonian \eqref{E:JaynesCummingsHamiltonian} decomposes into a sum over local contributions, the generating functional decomposes into products of purely local cumulants:
\begin{align}\label{E:DefinitionGeneratingFunctional}
 C_{0}^{(0)}[j,j^{*}]=&\underset{i}{\prod}\ln\left\langle\hat{T}e^{-\underset{0}{\overset{\beta}{\int}}d\tau\left[j_{i}(\tau)\hat{a}_{i}^{\dagger}(\tau)+j_{i}^{*}(\tau)\hat{a}_{i}(\tau)\right]}\right\rangle _{0}.
\end{align}
The local cumulants then follow from
\begin{align}\label{E:DefinitionHigherOrderCummulants}
 &C_{n}^{(0)}\left(i_{1}^{\prime},\tau_{1}^{\prime};\ldots;i_{n}^{\prime},\tau_{n}^{\prime}|i_{1},\tau_{1};\ldots;i_{n},\tau_{n}\right)\\
 &\;=\left.\frac{\delta^{2n}C_{0}^{(0)}\left[j,j^{*}\right]}{\delta j_{i_{1}^{\prime}}\left(\tau_{1}^{\prime}\right)\ldots\delta j_{i_{n}^{\prime}}\left(\tau_{n}^{\prime}\right)\delta j_{i_{1}}^{*}\left(\tau_{1}\right)\ldots\delta j_{i_{n}}^{*}\left(\tau_{n}\right)}\right|_{j=j^{*}=0}.\notag
\end{align}
Due to the local structure of \eqref{E:DefinitionGeneratingFunctional} all cumulants \eqref{E:DefinitionHigherOrderCummulants} vanish unless the site indexes are all equal which yields the relation
\begin{align}\label{E:DefinitionHigherOrderCummulantsLocality}
 &C_{n}^{(0)}\left(i_{1}^{\prime},\tau_{1}^{\prime};\ldots;i_{n}^{\prime},\tau_{n}^{\prime}|i_{1},\tau_{1};\ldots;i_{1},\tau_{n}\right)\notag\\
 &=C_{n}^{(0)}\left(i_{1};\tau_{1}^{\prime},\ldots,\tau_{n}^{\prime}|\tau_{1},\ldots,\tau_{n}\right)\underset{\alpha,\beta}{\prod}\delta_{i_{\alpha}^{\prime},i_{\beta}}.
\end{align}
Hence, we just have to calculate the \textit{local} cumulants $C_{n}^{(0)}\left(i_{1};\tau_{1}^{\prime},\ldots,\tau_{n}^{\prime}|\tau_{1},\ldots,\tau_{n}\right)$. Performing the calculations according to formula \eqref{E:DefinitionHigherOrderCummulants} and rearranging the resulting terms yields the cumulant decomposition for each thermal Green function \eqref{E:DefinitionOfNParticleGreensFunction}. The lowest order Green functions read
\begin{gather}
 G_{1}^{(0)}\left(i,\tau_{1}|j,\tau_{2}\right)=C_{1}^{(0)}\left(i;\tau_{1}|\tau_{2}\right)\delta_{ij},\notag\\
  G_{2}^{(0)}\left(i,\tau_{1};j,\tau_{2}|k,\tau_{3};l,\tau_{4}\right)=C_{2}^{(0)}\left(i;\tau_{1},\tau_{2}|\tau_{3},\tau_{4}\right)\delta_{ij}\delta_{jk}\delta_{kl}\notag\\
  +C_{1}^{(0)}\left(i;\tau_{1}|\tau_{3}\right)\,C_{1}^{(0)}\left(j;\tau_{2}|\tau_{4}\right)\delta_{ik}\,\delta_{jl}\notag\\
  +C_{1}^{(0)}\left(i;\tau_{1}|\tau_{4}\right)\,C_{1}^{(0)}\left(j;\tau_{2}|\tau_{3}\right)\delta_{il}\,\delta_{jk}.\label{E:GreensFunction40inCummulants}
\end{gather}
With this cumulant decomposition we find the following expansion of the free energy functional
\begin{widetext}
\begin{align}\label{E:FreeEnergyExpansionInGeneral}
 \mathcal{F}&\left[j,j^{*}\right]=\mathcal{F}_{0}-\frac{1}{\beta}\sum_{i,j}\int_{0}^{ \beta}d\tau_{1}\,\int_{0}^{ \beta}d\tau_{2}\,\Bigg\{ \bigg[a_{2}^{(0)}\left(i;\tau_{1}|\tau_{2}\right)\delta_{ij}+a_{2}^{(1)}\left(i;\tau_{1}|\tau_{2};j\right)\bigg]\,j_{i}(\tau_{1})\,j_{j}^{*}(\tau_{2})\notag\\
 &+\frac{1}{4}\int_{0}^{ \beta}d\tau_{3}\,\int_{0}^{ \beta}d\tau_{4}\bigg[a_{4}^{(0)}\left(i;\tau_{1},\tau_{3}|\tau_{2},\tau_{4}\right)\delta_{ij}+2\,a_{4}^{(1)}\left(i;\tau_{1},\tau_{3}|\tau_{2},\tau_{4};j\right)\bigg]\,j_{j}(\tau_{1})\,j_{i}(\tau_{3})\,j_{i}^{*}(\tau_{2})\,j_{j}^{*}(\tau_{4})\notag\\
 &+\frac{1}{2}\int_{0}^{ \beta}d\tau_{3}\,\int_{0}^{ \beta}d\tau_{4}\,\tilde{a}_{4}^{(1)}\left(i;\tau_{1},\tau_{3}|\tau_{2},\tau_{4};j\right)\,j_{j}(\tau_{1})\,j_{i}(\tau_{3})\,j_{i}^{*}(\tau_{2})\,j_{j}^{*}(\tau_{4})\Bigg\},
\end{align}
\end{widetext}
where the introduced expansion coefficients are defined as
\begin{align}\label{E:FreeEnergyExpansionCoefficients}
       &a_{2}^{(0)}\left(i;\tau_{1}|\tau_{2}\right)= C_{1}^{(0)}\left(i;\tau_{1}|\tau_{2}\right)\,,\\
       &a_{2}^{(1)}\left(i;\tau_{1}|\tau_{2};j\right)=\kappa_{ij}\int_{0}^{ \beta}d\tau\,C_{1}^{(0)}\left(i;\tau_{1}|\tau\right)\,\,C_{1}^{(0)}\left(j;\tau|\tau_{2}\right)\,,\notag\\
       &a_{4}^{(0)}\left(i;\tau_{1},\tau_{3}|\tau_{2},\tau_{4}\right)= C_{2}^{(0)}\left(i;\tau_{1},\tau_{3}|\tau_{2},\tau_{4}\right)\,,\notag\\
       &a_{4}^{(1)}\left(i;\tau_{1},\tau_{2}|\tau_{3},\tau_{4};j\right)=\kappa_{ij}\int_{0}^{ \beta}d\tau\,C_{2}^{(0)}\left(i;\tau,\tau_{2}|\tau_{3},\tau_{4}\right)\notag\\
       &\times C_{1}^{(0)}\left(j;\tau_{1}|\tau\right),\notag\\
       &\tilde{a}_{4}^{(1)}\left(i;\tau_{1},\tau_{3}|\tau_{2},\tau_{4};j\right)=\kappa_{ij}\int_{0}^{ \beta}d\tau\,C_{2}^{(0)}\left(i;\tau_{1},\tau_{2}|\tau_{3},\tau\right)\notag\\
       &\times\,C_{1}^{(0)}\left(j;\tau|\tau_{4}\right).\notag
\end{align}
Due to the locality of $\hat{H}_{0}$ in Eq.~\eqref{E:JaynesCummingsHamiltonian} the cumulants $C_{n}^{(0)}$ do not depend on the site indexes $i,j$. For this reason, we will drop the site index in the following calculations for convenience. Finally, we notice that the form of the above coefficients $a_{2}^{(1)}, a_{4}^{(1)}$, and $\tilde{a}_{4}^{(1)}$
can be further simplified by going into frequency space. Therefore we perform the Matsubara transformation
\begin{gather}
  f\left(\omega_{\rm{m}}\right)=\frac{1}{\sqrt{\beta}}\int_{0}^{\beta}d\tau \,f\left(\tau\right)e^{i\,\omega_{\rm{m}}\tau},\label{E:DefinitionMatsubaraTrans}\\
  f\left(\tau\right)=\frac{1}{\sqrt{\beta}}\sum_{m=-\infty}^{\infty}f\left(\omega_{\rm{m}}\right)e^{-i\,\omega_{\rm{m}}\tau},
\end{gather}
with the Matsubara frequencies
\begin{equation}\label{E:MatsubaraFrequency}
 \omega_{\rm{m}}=\frac{2\,\pi\,\rm{m}}{\beta},\hspace{1cm}\rm{m}\,\in\,\mathbb{Z}.
\end{equation}
At first, we calculate the coefficient $a_{2}^{(0)}\left(\omega_{\rm{m1}}|\omega_{\rm{m2}}\right)$ in Matsubara space. Due to frequency conservation the following relation has to hold
\begin{equation}
 a_{2}^{(0)}\left(\omega_{\rm{m1}}|\omega_{\rm{m2}}\right)=a_{2}^{(0)}\left(\omega_{\rm{m1}}\right)\,\delta_{\omega_{\rm{m1}},\omega_{\rm{m2}}}.
\end{equation}
This coefficient can be derived from the expression \eqref{E:FreeEnergyExpansionCoefficients} with the help of relations \eqref{E:DefinitionOfNParticleGreensFunction} and \eqref{E:GreensFunction40inCummulants} by performing a Matsubara transformation \eqref{E:DefinitionMatsubaraTrans}. Using the polariton mapping introduced in Ref.~\cite{key-2} to calculate the thermal expectation values, we obtain the following result:
\begin{align*}
 &a_{2}^{(0)}\left(\omega_{\rm{m1}}\right)=\frac{1}{\mathcal{Z}_{0}}\sum_{\alpha,\alpha^{\prime}=\pm}\Bigg\{ \frac{\left(t_{1\alpha^{\prime}-}\right)^{2}}{E_{1\alpha^{\prime}}-i\,\omega_{\rm{m}}}-\sum_{n=1}^{\infty}e^{-\beta E_{n\alpha}} \notag
\end{align*}
\begin{align}
  &\times\left[\frac{\left(t_{(n+1)\alpha^{\prime}\alpha}\right)^{2}}{E_{n\alpha}-E_{(n+1)\alpha^{\prime}}+i\,\omega_{\rm{m}}}-\frac{\left(t_{n\alpha\alpha^{\prime}}\right)^{2}}{E_{(n-1)\alpha^{\prime}}-E_{n\alpha}+i\,\omega_{\rm{m}}}\right]\Bigg\}.\label{E:A20inMatsubaraSpace}
\end{align}
The coefficients $t_{n\alpha\beta}$ in the above expression stem from the fact that there exist two kind of polariton species, where the lower branch is labeled by $\alpha,\beta=-1$ and the upper branch by $\alpha,\beta=+1$. These coefficients are defined as
\begin{subequations}\label{E:PolaritonMappingTransitionAmplitudes}
\begin{align}
 t_{n\pm-}=\sqrt{n}\,a_{n}^{\pm}\,b_{n-1}^{+}+\sqrt{n-1}\,b_{n}^{\pm}\,b_{n-1}^{-} \label{E:PolaritonMappingTransitionAmplitude-}\,,\\
 t_{n\pm+}=\sqrt{n}\,a_{n}^{\pm}\,a_{n-1}^{+}+\sqrt{n-1}\,b_{n}^{\pm}\,a_{n-1}^{-} \label{E:PolaritonMappingTransitionAmplitude+}\,,
\end{align}
\end{subequations}
with mixing angle dependent amplitudes given by
\begin{equation}
 a_{n}^{\alpha}=\left\{ \begin{array}{c}
\sin\theta_{n}\,,\;\alpha=+\\
\cos\theta_{n}\,,\;\alpha=-\end{array}\right.,\hspace{.1em} b_{n}^{\alpha}=\left\{ \begin{array}{c}
\cos\theta_{n}\,,\;\alpha=+\\
-\sin\theta_{n}\,,\;\alpha=-\,.\end{array}\right.
\end{equation}
With the help of this result we can also determine the higher hopping corrections. By using frequency conservation again we find the relation
\begin{align}\label{E:ConvolutionProperty}
 a_{2}^{(1)}\left(\omega_{\rm{m1}}|\omega_{\rm{m2}}\right)=&a_{2}^{(0)}\left(\omega_{\rm{m1}}\right)a_{2}^{(0)}\left(\omega_{\rm{m2}}\right)\delta_{\omega_{\rm{m1}},\omega_{\rm{m2}}}
\end{align}
and
\begin{align}\label{E:4orderMatsubaraCoefficient}
  a_{4}^{(0)}&\left(\omega_{\rm{m1}},\omega_{\rm{m3}}|\omega_{\rm{m2}},\omega_{\rm{m4}}\right)=\frac{1}{\beta^{2}} \;\delta_{\omega_{\rm{m1}}+\omega_{\rm{m3}},\omega_{\rm{m2}}+\omega_{\rm{m4}}}\notag\\
  &\left\lbrace -\,a_{2}^{(0)}\left(\omega_{\rm{m1}}\right)a_{2}^{(0)}\left(\omega_{\rm{m3}}\right)\left[\delta_{\omega_{\rm{m1}},\omega_{\rm{m2}}}\,\delta_{\omega_{\rm{m3}},\omega_{\rm{m4}}}\right.\right.\notag\\
  &\left. +\,\delta_{\omega_{\rm{m1}},\omega_{\rm{m4}}}\,\delta_{\omega_{\rm{m3}},\omega_{\rm{m2}}}\right]+\int_{0}^{\beta}d\tau_{1}\ldots d\tau_{4}\notag\\
  &\times\,\left\langle \hat{T}\left[\hat{a}^{\dagger}(\tau_{1})\,\hat{a}^{\dagger}(\tau_{3})\,\hat{a}(\tau_{2})\,\hat{a}(\tau_{4})\right]\right\rangle _{0}\notag\\
  &\left.\times\,e^{i\left(-\omega_{\rm{m1}}\tau_{1}+\omega_{\rm{m2}}\tau_{2}-\omega_{\rm{m3}}\tau_{3}+\omega_{\rm{m4}}\tau_{4}\right)}\right\rbrace.
\end{align}
The calculation of the latter expression is complicated and rather lengthy. Therefore, we put a detailed calculation of this quantity in the appendix. Nevertheless, from general considerations like frequency conservation and integral properties in Matsubara space, we can deduce right away that the first order hopping correction is of the form
\begin{align}
 a_{4}^{(1)}&\left(\omega_{\rm{m1}},\omega_{\rm{m3}}|\omega_{\rm{m2}},\omega_{\rm{m4}}\right)=a_{2}^{(0)}\left(\omega_{\rm{m2}}\right)\notag\\
  &\times a_{4}^{(0)}\left(\omega_{\rm{m1}},\omega_{\rm{m3}}|\omega_{\rm{m4}}\right)\delta_{\omega_{\rm{m1}}+\omega_{\rm{m3}},\omega_{\rm{m2}}+\omega_{\rm{m4}}}.
\end{align}

\subsection{Ginzburg-Landau Theory}
Within this section we finally derive the Ginzburg-Landau action for the Jaynes-Cummings-Hubbard model, which is the proper thermodynamic potential to describe the quantum phase transition of this system. Since the symmetry-breaking currents $j,j^*$ are no physical quantities, one has to transform them into physical fields in terms of the order parameter. This is accomplished by means of a Legendre transformation of the free-energy functional to an effective action as has already been explored in detail in the context of the thermal phase transitions \cite{Kleinert,Zinn}. In order to do this in a concise way, we first rewrite the grand-canonical free energy \eqref{E:FreeEnergyExpansionInGeneral} in the following form in Matsubara space
\begin{align}\label{E:FreeEnergyInMatsubara}
  &\mathcal{F}\left[j,j^{*}\right]=\mathcal{F}_{0}-\frac{1}{\beta}\sum_{i,j}\sum_{\omega_{\rm{m1}},\omega_{\rm{m2}}}\left\{ M_{ij}\left(\omega_{\rm{m1}},\omega_{\rm{m2}}\right)j_{i}\left(\omega_{\rm{m1}}\right)\right.\notag\\
  &\times\; j_{j}^{*}\left(\omega_{\rm{m2}}\right)+\sum_{k,l}\sum_{\omega_{\rm{m3}},\omega_{\rm{m4}}}N_{ijkl}\left(\omega_{\rm{m1}},\omega_{\rm{m2}},\omega_{\rm{m3}},\omega_{\rm{m4}}\right)\notag\\
  &\left.\times\; j_{i}(\omega_{\rm{m1}})j_{j}(\omega_{\rm{m3}})j_{k}^{*}(\omega_{\rm{m2}})j_{l}^{*}(\omega_{\rm{m4}})\right\}+\ldots\,,
\end{align}
where we have introduced the abbreviations
\begin{align}\label{E:DefinitionCoefficientM}
  M_{ij}(\omega_{\rm{m1}},&\omega_{\rm{m2}})=\left[a_{2}^{(0)}\left(\omega_{\rm{m1}}\right)\delta_{i,j}\right.\notag \\
  &\left.+\kappa_{ij}\,a_{2}^{(0)}\left(\omega_{\rm{m1}}\right)a_{2}^{(0)}\left(\omega_{\rm{m2}}\right)\right]\delta_{\omega_{\rm{m1}},\omega_{\rm{m2}}}\,,
\end{align}
and
\begin{align}\label{E:DefinitionCoefficientN}
  N_{ijkl}&\left(\omega_{\rm{m1}},\omega_{\rm{m2}},\omega_{\rm{m3}},\omega_{\rm{m4}}\right)=\frac{1}{4}a_{4}^{(0)}\left(\omega_{\rm{m1}},\omega_{\rm{m3}}|\omega_{\rm{m4}}\right)\notag\\
  &\times\left[\delta_{i,j}\,\delta_{k,l}\,\delta_{i,k}+\kappa_{ik}\,a_{2}^{(0)}\left(\omega_{\rm{m2}}\right)\delta_{i,j}\,\delta_{i,l}\right.\notag\\
  &+\left.\kappa_{ij}\,a_{2}^{(0)}\left(\omega_{\rm{m3}}\right)\delta_{i,k}\,\delta_{k,l}\right]\delta_{\omega_{\rm{m1}}+\omega_{\rm{m3}},\omega_{\rm{m2}}+\omega_{\rm{m4}}}\,.
\end{align}
Now we define the Legendre transformation by self-consistently introducing the order parameter field $\Psi_{i}(\omega_{\rm{m}})$ according to
\begin{equation}\label{E:GLorderParameter}
 \Psi_{i}\left(\omega_{\rm{m}}\right)=\left\langle \hat{a}_{i}(\omega_{\rm{m}})\right\rangle _{0}=\beta\frac{\delta\mathcal{F}}{\delta j_{i}^{*}\left(\omega_{\rm{m}}\right)}\,.
\end{equation}
Note, that this Ginzburg-Landau order parameter field differs from the Landau order parameter by being space \textit{and} time dependent. Inserting expression \eqref{E:FreeEnergyInMatsubara} into equation \eqref{E:GLorderParameter} yields the following relation for the Ginzburg-Landau order parameter field
\begin{align}\label{E:OrderFieldFunctionalOfCurrents}
 \Psi_{i}\left(\omega_{\rm{m}}\right)=&-\sum_{p}\sum_{\omega_{\rm{m1}}}\left\lbrace M_{pi}\left(\omega_{\rm{m1}},\omega_{\rm{m}}\right)\; j_{p}\left(\omega_{\rm{m1}}\right)\phantom{\sum_i}\right.\notag\\
 &-2\sum_{k,l}\sum_{\omega_{\rm{m2}},\omega_{\rm{m3}}}N_{lpki}\left(\omega_{\rm{m1}},\omega_{\rm{m2}},\omega_{\rm{m3}},\omega_{\rm{m}}\right)\notag\\
  &\left.\times\; j_{l}(\omega_{\rm{m1}})j_{p}(\omega_{\rm{m3}})j_{k}^{*}(\omega_{\rm{m2}})\right\rbrace+\ldots\,.
\end{align}
Furthermore, relation \eqref{E:GLorderParameter} motivates to introduce the effective action
\begin{align}\label{E:EffectiveActionDefinition}
 \Gamma\left[\Psi_{i}\left(\omega_{\rm{m}}\right),\Psi_{i}^{*}\left(\omega_{\rm{m}}\right)\right]=&\mathcal{F}\left[j,j^{*}\right]-\frac{1}{\beta}\sum_{i,\omega_{\rm{m}}}\left[\Psi_{i}\left(\omega_{\rm{m}}\right)j_{i}^{*}\left(\omega_{\rm{m}}\right)\right.\notag\\
  &\left. +\;\Psi_{i}^{*}\left(\omega_{\rm{m}}\right)j_{i}(\omega_{\rm{m}})\right]\,,
\end{align}
where $\Psi$ and $j$ are conjugate variables satisfying the Legendre relations
\begin{equation}
 j_{i}(\omega_{\rm{m}})=-\beta\frac{\delta\Gamma}{\delta\Psi_{i}^{*}\left(\omega_{\rm{m}}\right)}\,,\hspace{.5em}
 j^{*}_{i}(\omega_{\rm{m}})=-\beta\frac{\delta\Gamma}{\delta\Psi_{i}\left(\omega_{\rm{m}}\right)}\,.
\end{equation}
Using the fact, that physical situations correspond to vanishing currents $j=j^{*}=0$, yields the following \textit{equations of motion}
\begin{equation}\label{E:DefinitionPhysicalEOM}
 \frac{\delta\Gamma}{\delta\Psi_{i}^{*}\left(\omega_{\rm{m}}\right)}=\frac{\delta\Gamma}{\delta\Psi_{i}\left(\omega_{\rm{m}}\right)}=0\,.
\end{equation}
Thus, the effective action is stationary with respect to the order parameter field. Now, in order to determine the explicit form of the effective action, we need to express all symmetry-breaking currents $j$ by the Ginzburg-Landau order parameter field $\Psi$. Therefore, we recursively invert relation \eqref{E:OrderFieldFunctionalOfCurrents} up to first order in the hopping strength $\kappa$, which yields
\begin{align}\label{E:JasFunctionalOfOnlyPsi}
 j_{i}\left(\omega_{\rm{m}}\right)=&-\sum_{p}\sum_{\omega_{\rm{m1}}}M_{ip}^{-1}\left(\omega_{\rm{m}},\omega_{\rm{m1}}\right)\left[\Psi_{p}\left(\omega_{\rm{m1}}\right)\phantom{\sum_i}\right.\notag\\
 &-2\sum_{q,k,l}\sum_{\omega_{\rm{m2}},\omega_{\rm{m3}}}N_{lqkp}\left(\omega_{\rm{m1}},\omega_{\rm{m2}},\omega_{\rm{m3}},\omega_{\rm{m}}\right)\notag\\
  &\left.\times\;J_{l}(\omega_{\rm{m1}})J_{q}(\omega_{\rm{m3}})J_{k}^{*}(\omega_{\rm{m2}})\right]+\ldots\,,
\end{align}
where we define the abbreviations
\begin{equation}\label{E:DefinitionJ}
 J_{i}\left(\omega_{\rm{m}}\right)=-\sum_{p}\sum_{\omega_{\rm{m1}}}M_{ip}^{-1}\left(\omega_{\rm{m1}},\omega_{\rm{m}}\right)\Psi_{p}\left(\omega_{\rm{m1}}\right)\,,
\end{equation}
and
\begin{equation}\label{E:MinvFromA20}
 M_{ij}^{-1}\left(\omega_{\rm{m1}},\omega_{\rm{m2}}\right)=\frac{\delta_{\omega_{\rm{m1}},\omega_{\rm{m2}}}}{a_{2}^{(0)}\left(j,\omega_{\rm{m1}}\right)}\left[\delta_{i,j}-a_{2}^{(0)}\left(j,\omega_{\rm{m2}}\right)\kappa_{ij}\right].
\end{equation}
Inserting the above relations into the Legendre transformation \eqref{E:EffectiveActionDefinition} gives the explicit expression for the effective action up to the desired order in the hopping strength
\begin{align}\label{E:GLeffectiveAction}
 &\Gamma\left[\Psi_{i}\left(\omega_{\rm{m}}\right),\Psi_{i}^{*}\left(\omega_{\rm{m}}\right)\right]= \mathcal{F}_{0}+\frac{1}{\beta}\sum_{i}\sum_{\omega_{\rm{m1}}}\left\{ \frac{\left|\Psi_{i}\left(\omega_{\rm{m1}}\right)\right|^{2}}{a_{2}^{(0)}\left(\omega_{\rm{m1}}\right)}\right.\notag\\
  &-\sum_{j}\kappa_{ij}\,\Psi_{j}\left(\omega_{\rm{m1}}\right)\Psi_{i}^{*}\left(\omega_{\rm{m1}}\right)-\sum_{\omega_{\rm{m2}},\omega_{\rm{m3}},\atop\omega_{\rm{m4}}}\Psi_{i}\left(\omega_{\rm{m1}}\right)\Psi_{i}^{*}\left(\omega_{\rm{m2}}\right)\notag\\
 &\times\left.\frac{a_{4}^{(0)}\left(\omega_{\rm{m1}},\omega_{\rm{m3}}|\omega_{\rm{m2}},\omega_{\rm{m4}}\right)\Psi_{i}\left(\omega_{\rm{m3}}\right)\Psi_{i}^{*}\left(\omega_{\rm{m4}}\right)}{4\,a_{2}^{(0)}\left(\omega_{\rm{m1}}\right)a_{2}^{(0)}\left(\omega_{\rm{m2}}\right)a_{2}^{(0)}\left(\omega_{\rm{m3}}\right)a_{2}^{(0)}\left(\omega_{\rm{m4}}\right)}\right\}\notag\\
 &+\,\ldots.\;.
\end{align}
Equation \eqref{E:GLeffectiveAction} is the thermodynamic potential for the Jaynes-Cummings Hubbard model up to the desired accuracy in both the order parameter and the hopping parameter. However, as should be clear from the approximations performed so far, this expression can, in principle, be extended to include higher order corrections. In the next section this result is used to analyze several properties of the considered system.

\section{Results}
Having derived the effective action in the previous chapter, we now use this result to extract both thermodynamic and dynamic properties of the Jaynes-Cummings-Hubbard model. The starting point for this analysis is the equation of motion \eqref{E:DefinitionPhysicalEOM}. Inserting \eqref{E:GLeffectiveAction} yields
\begin{align}\label{E:ExplicitEOM}
 0=&\left[\frac{1}{a_{2}^{(0)}\left(\omega_{\rm{m}}\right)}-\sum_{j}\kappa_{ij}\right]\Psi_{i}\left(\omega_{\rm{m}}\right)-\sum_{\omega_{\rm{m1}},\omega_{\rm{m2}},\omega_{\rm{m3}}}\\
 \times&\frac{a_{4}^{(0)}\left(\omega_{\rm{m1}},\omega_{\rm{m3}}|\omega_{\rm{m2}},\omega_{\rm{m}}\right)\Psi_{i}\left(\omega_{\rm{m1}}\right)\,\Psi_{i}\left(\omega_{\rm{m3}}\right)\,\Psi_{i}^{*}\left(\omega_{\rm{m2}}\right)}{2\, a_{2}^{(0)}\left(\omega_{\rm{m1}}\right)a_{2}^{(0)}\left(\omega_{\rm{m2}}\right)a_{2}^{(0)}\left(\omega_{\rm{m3}}\right)a_{2}^{(0)}\left(\omega_{\rm{m}}\right)}.\notag
\end{align}
Using a particular ansatz for the order parameter field allows to examine both static and dynamic order parameter fields $\Psi_{i}\left( \omega_{\rm{m}}\right)$.

\subsection{Static Results}
\begin{figure}
  \centering
  \hspace{-1em}
  \includegraphics[width=0.48\textwidth]{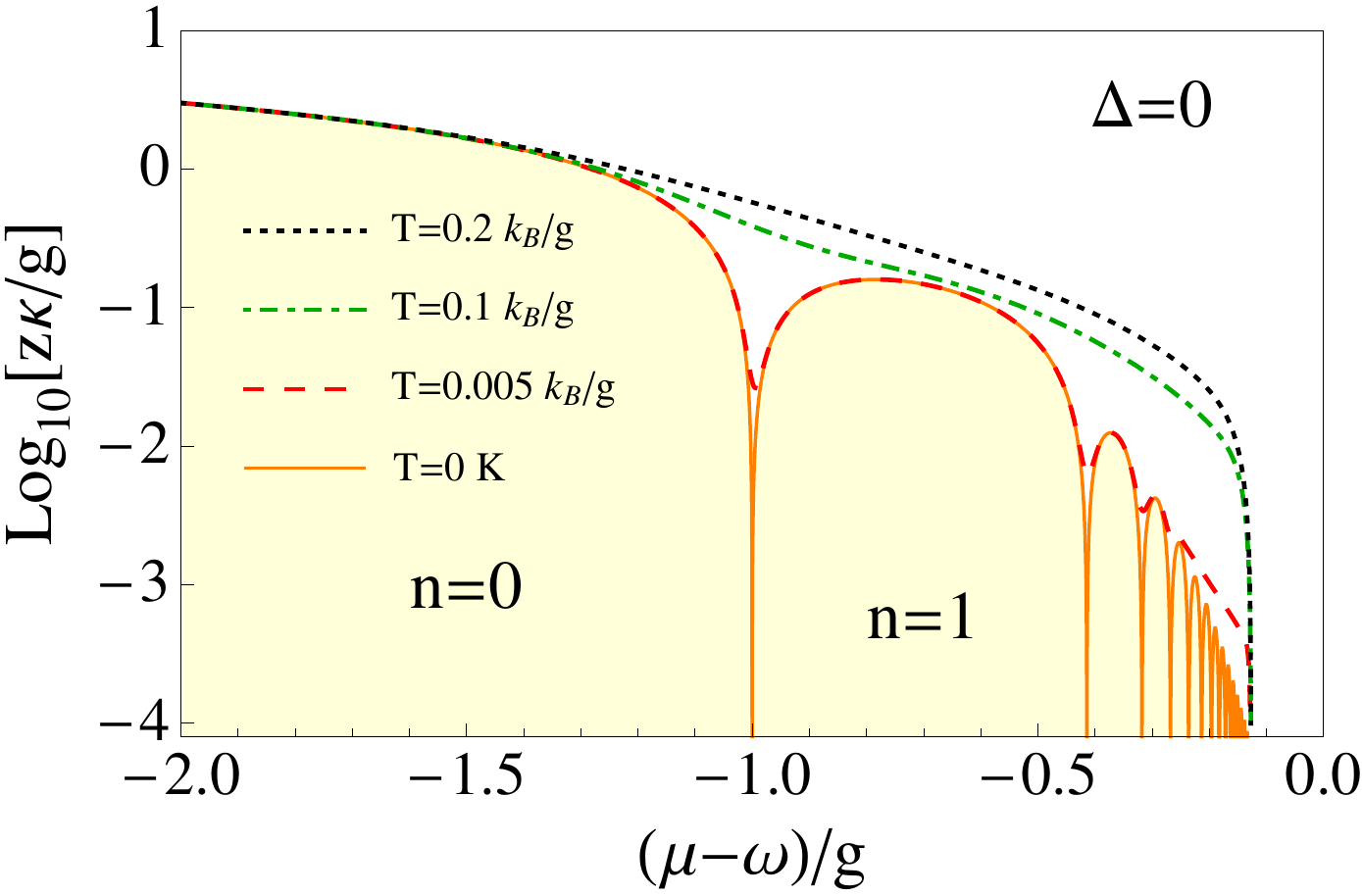}
  \caption[Quantum phase boundary for finite temperatures.]{(Color online) Quantum phase boundary for zero and finite temperatures at resonance $\Delta=0$. Interior of the lobes for $T=0K$ corresponds to the Mott insulator phase whereas the exterior corresponds to the superfluid phase. For finite temperatures the Mott insulator vanishes and becomes a mixture of normal and Mott phase. Note that the drop of the phase boundary at $\mu_{\rm{eff}}\approx-0.1g$ is a numerical remnant resulting from a summation cutoff in Eq.~\eqref{E:A20inMatsubaraSpace}. A full summation would lead to an infinite sequence of Mott lobes approaching $\mu_{\rm{eff}}=0$.}
  \label{F:QuantumPhaseBoundary}
\end{figure}

First, we consider an equilibrium situation, where the order parameter field is constant in both space and time
\begin{equation}\label{E:StaticFieldAnsatz}
 \Psi_{i}\left( \omega_{\rm{m}}\right) = \sqrt{\beta}\,\Psi^{\rm{eq}}_{i}\,\delta_{\omega_{\rm{m}},0}\,.
\end{equation}
Inserting this ansatz in the stationarity condition \eqref{E:ExplicitEOM} yields the following relation for the equilibrium order parameter
\begin{equation}\label{E:EquilibriumOrderParameter}
 \left|\Psi^{\text{eq}}\right|^{2}=\frac{2}{\beta}\,\frac{\left[a_{2}^{(0)}\left(0\right)\right]^{3}}{a_{4}^{(0)}\left(0,0|0,0\right)}\left[1-a_{2}^{(0)}\left(0\right)\,\kappa\, z\right]\,,
\end{equation}
where $z=2d$ represents the coordination number of the $d$-dimensional cubic lattice. Since the order parameter field is zero in the Mott insulator phase and takes on finite values in the superfluid regime, we can extract the quantum phase boundary from the condition that the equilibrium order parameter \eqref{E:EquilibriumOrderParameter} has to vanish. Thus the quantum phase boundary is defined via the relation
\begin{equation}\label{E:GLeqPhaseBoundary}
 \kappa\,z=\frac{1}{a_{2}^{(0)}\left(0\right)}\,.
\end{equation}
Together with the result \eqref{E:A20inMatsubaraSpace} this equation yields a phase diagram which is pictured in Fig.~\ref{F:QuantumPhaseBoundary} for vanishing detuning $\Delta=0$. Here we plot the effective hopping strength versus the effective chemical potential leading to a lobe structure where each lobe is associated to a specific mean on-site polariton number. The regions within these lobes correspond to the Mott insulator phase, whereas the exterior region corresponds to the superfluid regime. First we note that the phase boundary for zero temperature is consistent with the results from Refs.~\cite{key-2,key-3}. Moreover, our Ginzburg-Landau theory also yields the phase boundary for finite temperatures. From Fig.~\ref{F:QuantumPhaseBoundary} we can see that increasing the temperature leads to a smeared out phase boundary. Thermal fluctuations mostly affect the region between two neighboring Mott lobes, whereas the middle of each Mott lobe almost does not change. This effect is stronger for lobes with higher polariton number as these system configurations are rather unstable.
Since a genuine Mott insulator is defined by a vanishing compressibility $\kappa_{T}=-\frac{1}{N_s}\left.\frac{\partial^{2}\Gamma}{\partial\mu^{2}}\right|_{\Psi=\Psi_{\text{eq}}}$ we find that the Mott insulator phase is only present for zero temperature. For finite temperature it becomes a mixture of normal and Mott insulating phase instead \cite{Fleischauer}.

Note that, in order to validate our results obtained so far, we follow Ref.~\cite{key-1} and explicitly compare our expressions for $a_{2}^{(0)},a_{4}^{(0)}$ with a corresponding mean-field calculation. We get back the mean-field free energy $\mathcal{F}_{\rm{MF}}=\mathcal{F}_{0}-\sum_{i} \left( a_{2}^{\rm{MF}}\,|\Psi_{i}|^{2}+\frac{1}{4}\,\beta\,a_{4}^{\rm{MF}}|\Psi_i|^4\right)$ by formally identifying $j_{i}(\tau)=-\kappa\,z\,\Psi_i$, $a_{2}^{\rm{MF}}=\,a_{2}^{(0)}(0)\,(\kappa\,z)^{2}-\kappa\,z$ and $a_{4}^{\rm{MF}}=\,a_{4}^{(0)}(0,0|0,0)\,(\kappa\,z)^{4}$. With this we reproduce the mean-field results of Refs.~\cite{key-2,key-3} including the dependence of the quantum phase boundary on the detuning parameter $\Delta$. If the system is tuned out of resonance, i.e. $\Delta\neq0$, all Mott lobes with higher mean particle number than one shrink in size and are shifted to smaller values of the effective chemical potential $\mu_{\rm{eff}}$ irrespective of the sign of the detuning. The Mott-lobe with mean particle number $1$ grows in size for negative detuning and shrinks for positive detuning, whereas the Mott-lobe with mean particle number $0$ covers the rest of the phase diagram and, thus, is the only Mott insulator region that does not form a closed lobe. This special behavior of the first two Mott regions stems from the composite nature of the polaritons and has no analogue in the Bose-Hubbard model.
\subsection{Dynamic Results}
\begin{figure}[t]
 \centering{\hspace{-.2cm}\includegraphics[width=.4\textwidth]{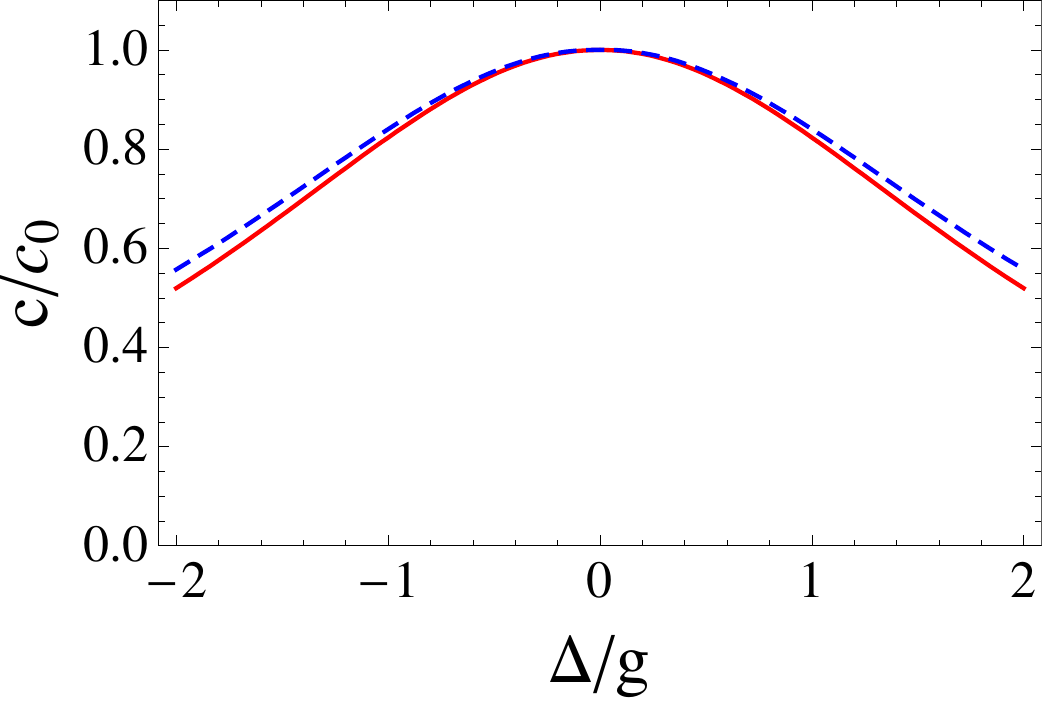}}
 \caption{(Color online) Sound velocity $c$ normalized by the sound velocity at zero detuning $c_0$ in dependence of the detuning parameter $\Delta$ for $n=2$ with $\kappa=0.001\kappa_{\rm{crit}}$ (solid) and $\kappa=0.15\kappa_{\rm{crit}}$ (dashed).}
 \label{F:subSoundD}
\end{figure}

Within this section we analyze the dynamic behavior of the JCH model. We especially focus on signatures within excitation spectra such as energy gaps, effective masses and the sound velocity of polaritons. These system properties are experimentally accessible via emission and transmission spectroscopy \cite{Ciuti2009}.In order to derive these properties we investigate the dynamic behavior of the effective action around the equilibrium fields. Thus, introducing the vector $\mathbf{\Psi}=(\Psi,\Psi^{*})$ we insert the ansatz $\mathbf{\Psi}=\mathbf{\Psi}_{\rm{eq}}+\delta\mathbf{\Psi}$ in the equation of motion \eqref{E:ExplicitEOM} and its conjugate complex. This yields a system of two coupled equations corresponding to the equations of motion for the elongations $\delta\mathbf{\Psi}$ around the order field. In order to find a non-trivial solution of the equation of motion, we obtain the relation
\begin{align}\label{E:ForthOrderEoM}
 0&\overset{!}{=}\left[ \frac{\delta^{2}\Gamma}{\delta\Psi_{i}\left( \omega_{\rm{m}}\right) \delta\Psi_{j}\left( -\omega_{\rm{m}}\right)} \frac{\delta^{2}\Gamma}{\delta\Psi_{i}^*\left( -\omega_{\rm{m}}\right) \delta\Psi_{j}^*\left( \omega_{\rm{m}}\right)} \right. \\
 &\left. -\frac{\delta^{2}\Gamma}{\delta\Psi_{i}\left( \omega_{\rm{m}}\right) \delta\Psi_{j}^*\left( \omega_{\rm{m}}\right)}\frac{\delta^{2}\Gamma}{\delta\Psi_{i}^*\left( -\omega_{\rm{m}}\right) \delta\Psi_{j}\left( -\omega_{\rm{m}}\right)}\right]_{\mathbf{\Psi}=\mathbf{\Psi}_{\rm{eq}}}\notag.
\end{align}
Inserting Eq.~\eqref{E:GLeffectiveAction} we see that, due to the effective hopping amplitude $\kappa_{ij}$, the second partial derivatives of the action still depend on the site distance $i-j$. This suggests to further apply a spatial Fourier transformation in order to simplify the calculations. Additionally, switching from Matsubara frequencies to continuous frequencies within a Wick rotation, yields the following explicit expression for Eq.~\eqref{E:ForthOrderEoM}
\begin{align}\label{E:ForthOrderEoMexplicit}
 0&\overset{!}{=}A(-\omega,\mathbf{k})\,A^*(\omega,\mathbf{k})-B^*(-\omega,\mathbf{k})\,B(\omega,\mathbf{k}),
\end{align}
with the abbreviations
\begin{align}
 A(\omega,\mathbf{k})=&\frac{1}{a_2^{(0)}(\mathbf{k},\omega)}-J(\mathbf{k})-\frac{a_4^{(0)}(\mathbf{k};\omega,0|0,\omega)|\Psi_{\rm{eq}}|^2}{\left[ a_2^{(0)}(\mathbf{k},\omega)\right]^2 \left[ a_2^{(0)}(\mathbf{k},0)\right]^2},
\end{align}
and
\begin{align}
 B(\omega,\mathbf{k})=&-\frac{a_4^{(0)}(\mathbf{k};0,0|-\omega,\omega)|\Psi_{\rm{eq}}|^2}{2\left[ a_2^{(0)}(\mathbf{k},0)\right]^2 a_2^{(0)}(\mathbf{k},\omega)a_2^{(0)}(\mathbf{k},-\omega)}.
\end{align}
Assuming a simple three-dimensional cubic lattice with lattice constant $a$ yields $J(\mathbf{k})=2\kappa\sum_{i=1}^3\cos\left( k_i a\right) $. Taking a closer look at Eq.~\eqref{E:ForthOrderEoMexplicit} we see that it implicitly defines the dispersion relation $\omega\left( \mathbf{k}\right) $. However, due to the complex expression found for $a_4^{(0)}$, a full evaluation of the above equations can only be done numerically. Just in the Mott insulator regime, where the $4$-point correlation $a_{4}^{(0)}$ drops out due to $\Psi_{\rm{eq}}=0$, it is possible to derive analytic expressions for the dispersions. In this special case we have to solve $1\overset{!}{=}a_2^{(0)}(\mathbf{k},\omega)\,J(\mathbf{k})$ leading to the dispersion relations
\begin{align}\label{E:DispRelW}
  \omega_{\pm}&(\mathbf{k})=\frac{1}{2}\bigg[E_{(n+1)-}-E_{(n-1)-}+J(\mathbf{k})\left(t_{n--}^{2}-t_{(n+1)--}^{2}\right)\notag\\
  &\pm\left(\left\{ E_{(n-1)-}-E_{(n+1)-}-J(\mathbf{k})\left[t_{n--}^{2}\pm t_{(n+1)--}^{2}\right]\right\} ^{2}\right.\notag\\
  &-4\Bigl\{ J(\mathbf{k})\, E_{(n+1)-}t_{n--}^{2}-E_{n-}^{2}+E_{(n-1)-}\notag\\
  &\times\left(E_{n-}-E_{(n+1)-}+J(\mathbf{k})\, t_{(n+1)--}^{2}\right)+E_{n-}\Bigl[E_{(n+1)-}\notag\\
  &\left.\left.-J(\mathbf{k})\left(t_{n--}^{2}+t_{(n+1)--}^{2}\right)\Bigr]\Bigr\} ^{\frac{1}{2}}\right)\right].
\end{align}
\begin{figure}[t]
  \centering \includegraphics[width=.48\textwidth]{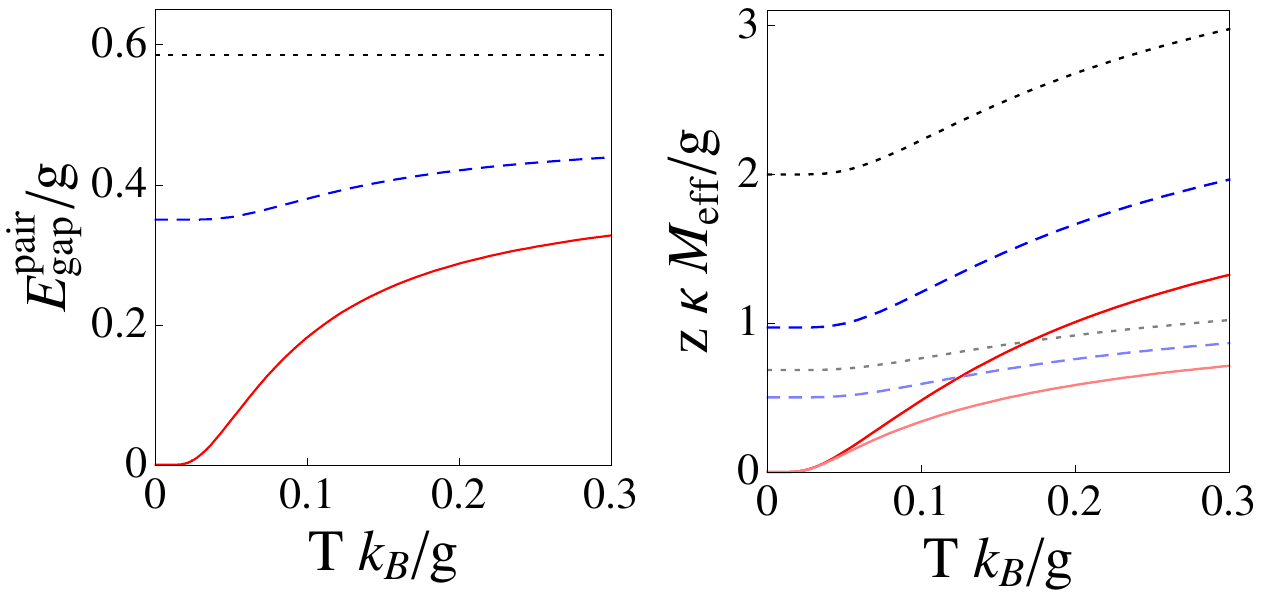}
  \caption{(Color online) The pair excitation energy gap (left ) and the effective mass (right) in dependence of the temperature for the first lobe $n=1$ at  $\mu_{\rm{eff}}=\mu_{\rm{crit}}=0.78g$ with zero detuning $\Delta=0$ and $\kappa=\kappa_{\rm{crit}}=0.16g$ (solid), $\kappa=0.1g$ (dashed) and $\kappa=10^{-4}\,g$ (dotted). The right picture represents the effective mass of both the hole (lower branch/light) and particle (upper branch/dark) excitations.}\label{F:subGapT}
\end{figure}
In order to clarify the physical meaning of equation \eqref{E:DispRelW} we set $J(\mathbf{k})=0$, which yields the simple relations
\begin{gather}
 \omega_{\rm{h}--}=E_{n-}-E_{(n-1)-}\,, \\
 \omega_{\rm{p}--}=E_{(n+1)-}-E_{n-}\,.
\end{gather}
\begin{figure*}
   \subfigure[\,Particle (dotted) and hole (dashed) dispersion relations in the Mott phase (gray), on the phase boundary (black) and in the superfluid phase (solid) in the direction $\mathbf{k}=k(1,1,1)$.]{\centering\includegraphics[width=0.8\textwidth]{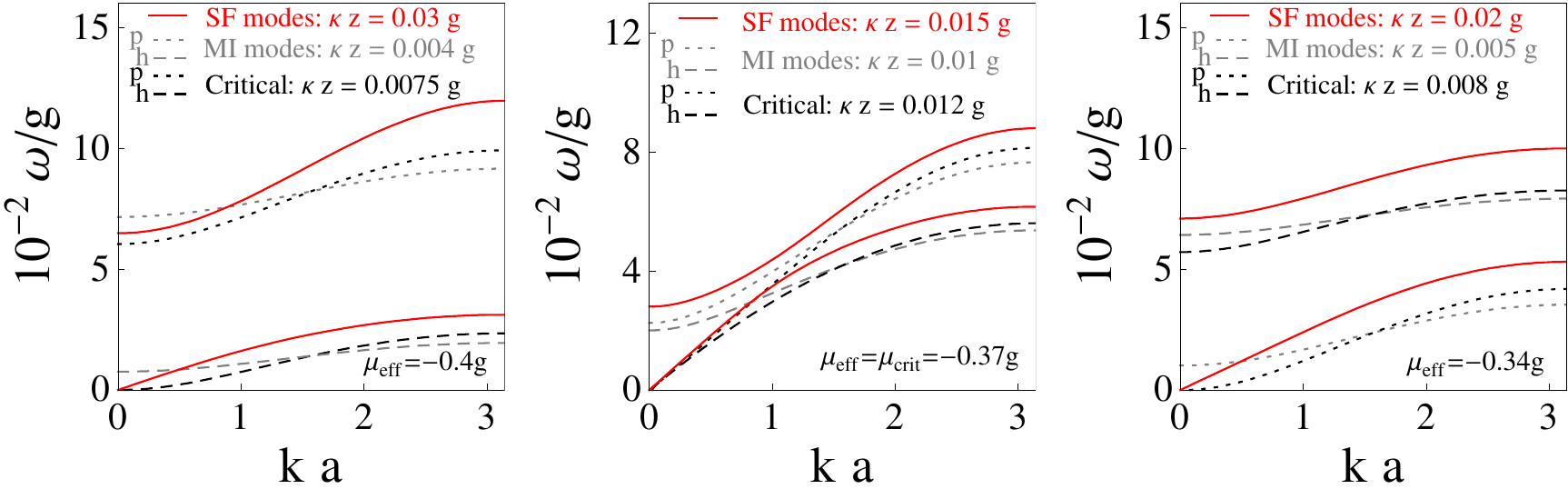}\label{F:subSpec}}
   \subfigure[\,Energy gap for particle (dotted) and hole (dashed) excitations in the Mott phase and for the massive mode in the superfluid phase (solid).]{\centering\includegraphics[width=0.8\textwidth]{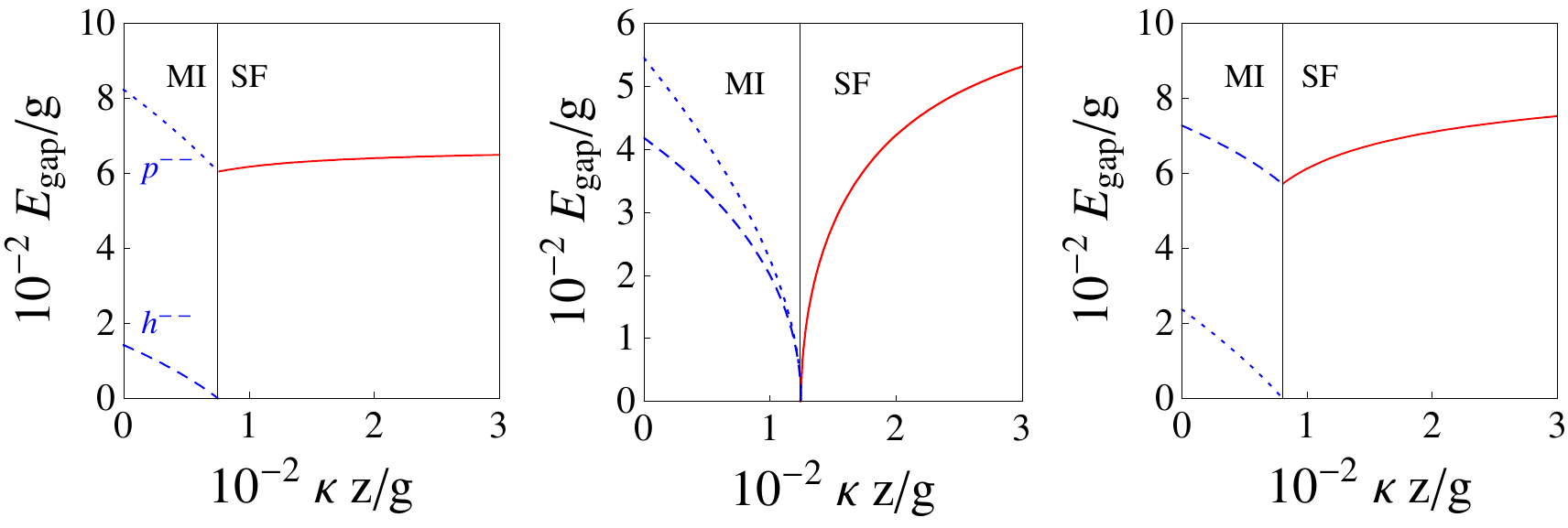}\label{F:subGap}}
   \subfigure[\,Effective mass for particle (dotted) and hole (dashed) excitations in the Mott phase and for the massive mode in the superfluid phase (solid).]{\centering\includegraphics[width=0.8\textwidth]{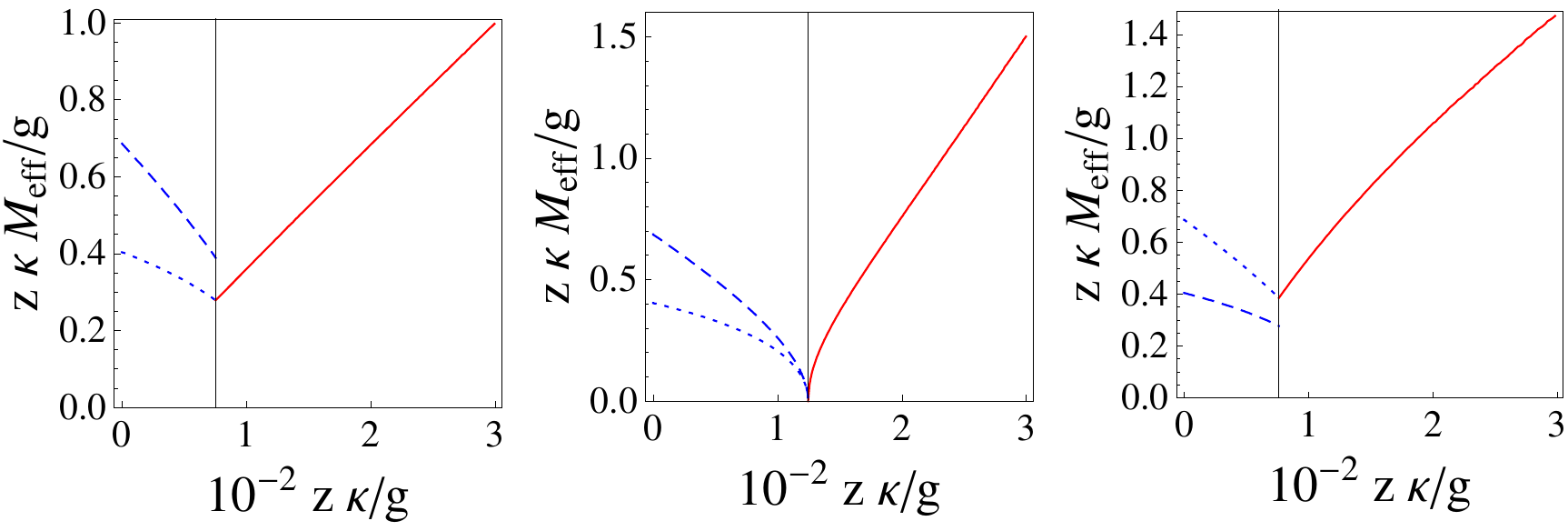}
   \label{F:subMass}}
   \subfigure[\,Sound velocity for the massive mode in the superfluid phase.]{\centering\includegraphics[width=0.8\textwidth]{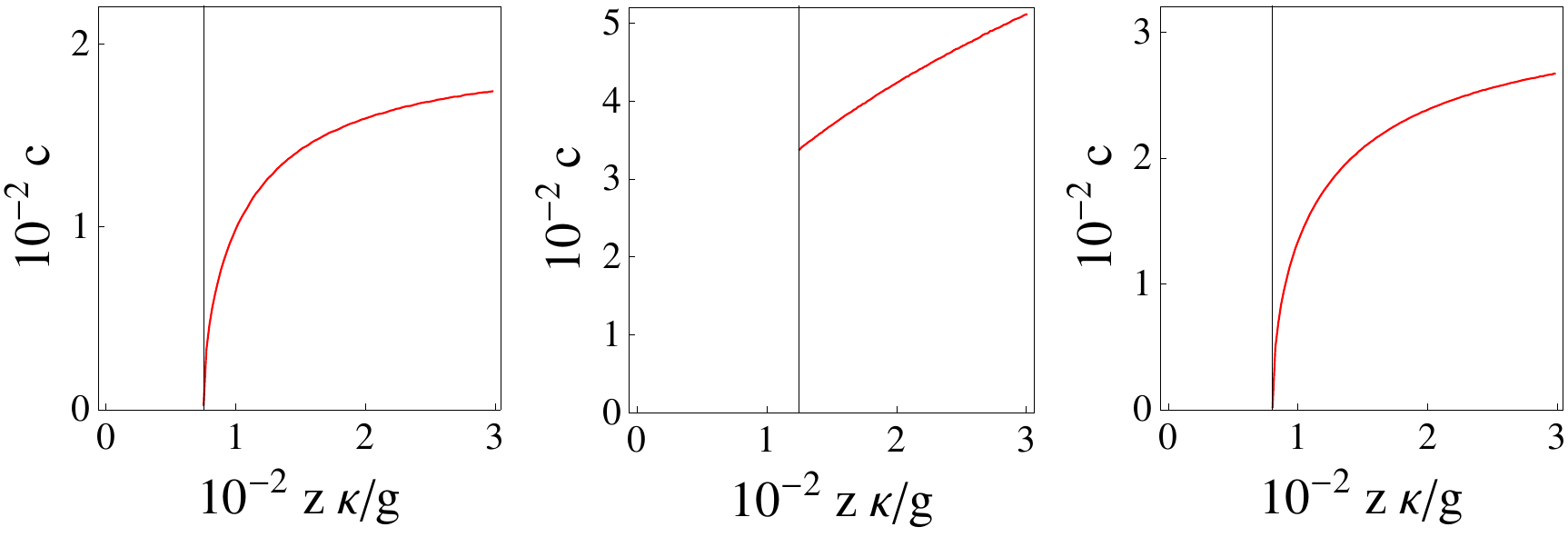}\label{F:subSoundK}}
   \caption{(Color online) Various dynamic results for a mean particle density $n=2$ at zero temperature and vanishing detuning $\Delta=0$. Left column at $\mu_{\rm{eff}}=-0.4g$, middle column for tip of the lobe at $\mu_{\rm{eff}}=-0.37g$ and right column at $\mu_{\rm{eff}}=-0.34g$.}\label{F:DynamicResults}
\end{figure*}
From these relations it is clear that $\omega_{\rm{h}--}$ ($\omega_{\rm{p}--}$) is the energy needed to remove (add) a lower-branch polariton from a lattice site, which is occupied by $n$ lower-branch polaritons. Therefore, we refer to these energies as the lower-branch-hole ($h$) and lower-branch-particle excitation ($p$), respectively. The same analysis can be applied to the full expression \eqref{E:ForthOrderEoMexplicit} leading to additional dispersion relation for upper-branch excitations $\omega_{\rm{h}++}$ and $\omega_{\rm{p}++}$ as well as mixed excitations  $\omega_{\rm{h}+-}$, $\omega_{\rm{h}-+}$, $\omega_{\rm{p}+-}$ and $\omega_{\rm{p}-+}$. However, these additional excitation channels are hardly of any interest for our considerations because they occur at much higher energies compared to the lower-branch polariton excitations. The lowest dispersion relation involving upper polariton states is $\omega_{\rm{h}+-}$, which lies at an energy around $2g$ for $n=2$. Therefore, we will just focus in all further calculations on the lower-branch excitations as they completely determine the low-temperature physics of the system. In order to extract more detailed quantities of interest, we subsequently expand the dispersion relations as follows
\begin{equation}
 \omega_{\rm{p,h}}(\mathbf{k})\thickapprox E_{\rm{gap}}+\frac{\mathbf{k}^{2}}{2\,M_{\rm{p,h}}}+\mathcal{O}(\mathbf{k}^4)\,.
\end{equation}
From this expansion we are able to derive, within the Mott insulator phase, the energy gap $E_{\rm{gap}}$ and the effective mass of the particle and hole excitations $M_{\rm{p,h}}$, respectively. As an example we evaluated their temperature dependence for the first Mott lobe $n=1$ with $\mu_{\rm{crit}}$ and zero detuning in Fig.~\ref{F:subGapT}. We find that, both the pair excitation energy gap as well as the effective mass of particle and hole excitations increase with higher temperatures. This effect becomes stronger as one approaches the critical hopping strength at the tip of the lobe.\\
Moreover, we expect to find in the superfluid regime, apart from the gapped mode, also a linear excitation mode which is associated with the broken symmetry in the superfluid regime according to the Nambu-Goldstone theorem \cite{nambu}. The dispersion relation for this mode reads
\begin{equation}
 \omega_{\rm{p,h}}(\mathbf{k})\thickapprox c\,|\mathbf{k}|+\mathcal{O}(\mathbf{k}^2).
\end{equation}
The results obtained from a numerical evaluation of the above formulas is presented in Fig.~\ref{F:DynamicResults} for a mean particle density $n=2$ at zero temperature and vanishing detuning $\Delta=0$. The pictures are arranged in a table in such a way that each line represents a specific physical quantity, for instance the second line  Fig.~\ref{F:subGap} shows plots for the energy gap. Furthermore, the plots in each column correspond to a fixed effective chemical potential. The left column shows plots for an effective chemical potential below the critical one, the middle column represents the critical effective chemical potential, and the right column corresponds to an effective chemical potential above the critical one. In the first line in Fig.~\ref{F:subSpec} we show the particle and hole excitation spectra in $\mathbf{k}=k(1,1,1)$ direction. In the Mott phase and at the phase boundary we always observe two excitation modes corresponding to the particle (dotted) and hole (dashed) excitations, respectively. These modes are always gapped in the Mott-insulator phase. By approaching the phase boundary at least one gap vanishes and thus for $\mu_{\rm{eff}}>\mu_{\rm{crit}}$ the particle mode becomes gapless at the phase border, whereas for $\mu_{\rm{eff}}<\mu_{\rm{crit}}$ the hole mode becomes gapless at the phase border. When approaching the phase boundary exactly at the lobe tip both particle and hole modes become gapless. Going further into the superfluid regime we find a gapped excitation mode as well as the anticipated gapless linear mode. In Fig.~\ref{F:subMass} we plot the corresponding effective masses. In the Mott phase we observe the masses of both the particle (dotted) and hole (dashed) excitations, whereas in the superfluid phase also only one massive mode survives. Additionally, we depict the sound velocity of the polariton excitations in the superfluid phase in Figs.~\ref{F:subSoundK} and \ref{F:subSoundD}. Figure \ref{F:subSoundK} shows the dependence of the sound velocity on the hopping strength. We find that it approaches a finite value at the tip of the Mott lobe, but vanishes at all other points of the Mott lobe. This behaviour shows that the JCH model has a dynamical critical exponent of $z=1$ which has been recently confirmed in a large-scale quantum Monte-Carlo simulation by M. Hohenadler et al.~\cite{Hohenadler2011}. Entering the superfluid phase the sound velocity increases steadily. However, as pictured in Fig.~\ref{F:subSoundD}, if the system is tuned out of resonance, i.e. $\Delta\neq0$, the sound velocity drops significantly. Finally, we note that our results Figs.~\ref{F:DynamicResults} and \ref{F:subSoundD} are in good qualitative agreement with the results from Ref.~\cite{Schmidt2010}. However, due to the restriction to the lowest hopping order in the effective action, our results lose validity deep in the superfluid phase. For this reason we do not obtain a shift of the maximum of the sound velocity in Fig.~\ref{F:subSoundD}, as is  observed in Ref.~\cite{Schmidt2010}. In order to obtain better results in the superfluid regime higher hopping corrections must be considered. Corresponding perturbative approaches for higher order corrections for the JCH model have been numerically calculated in Ref.~\cite{Hei2011}.
\section{Conclusion}
In summary, we successfully applied the cumulant expansion approach from Ref.~\cite{key-1} to derive a Ginzburg-Landau theory for the Jaynes-Cummings-Hubbard model up to fourth order in the symmetry-breaking currents and up to first order in the hopping strength. From the resulting effective action we extracted the phase diagram of the inherent quantum phase transition of the JCH model for finite temperature. In the case of vanishing temperature our result is in accordance with the results found from mean-field calculations \cite{key-2,key-3}. Subsequently, we derived the excitation spectra, energy gaps, and effective masses of the lower-branch polariton-excitations in the Mott insulator phase as well as in the superfluid phase. We investigated the temperature dependence of both the pair excitation energy gap and the effective mass of the particle and hole excitations in the Mott phase. Furthermore, we analyzed how the sound velocity in the superfluid phase depends at zero temperature on the hopping parameter and the detuning parameter. Finally, we point out that the the Ginzburg-Landau approach of this paper can be generalized to describe the real-time dynamics of the JCHM. This has already been shown for the real-time dynamics of the Bose-Hubbard model in Refs.~\cite{Grass2011,Grass2011a}. \\
We thank N. G. Berloff, C. Ciuti, and S. Schmidt for useful discussions and especially M. Hayn for valuable suggestions.

\appendix
\section{Fourth Order Coefficient}
Here we evaluate the fourth order coefficient \eqref{E:4orderMatsubaraCoefficient}, which involves the expectation values of time ordered products of four operators. First we notice that, for the time-ordered product of two annihilation and two creation operators, there are $6$ distinct permutations leading to different expectation values. Each of these orderings itself has $4$ time variable permutations corresponding to $\tau_{1}\leftrightarrow\tau_{2}$ and $\tau_{3}\leftrightarrow\tau_{4}$. Thus, overall one finds $24$ terms for the expectation value. Luckily, the integrals over different time-variable permutations yield the same result, and thus, they just lead to a fixed pre-factor $4$. For this reason, one just needs to determine the $6$ different thermal averages for one specific time-ordering. Furthermore, these expectation values are local quantities and, therefore, we drop the site indexes in the following calculations. Thus, one has to determine the following expressions:
\begin{align}\label{E:AllSix4PointCorrs}
 \left\langle \hat{a}^{\dagger}(\tau_{1})\,\hat{a}^{\dagger}(\tau_{3})\,\hat{a}(\tau_{2})\,\hat{a}(\tau_{4})\right\rangle _{0}\,,&\left\langle \hat{a}^{\dagger}(\tau_{1})\,\hat{a}(\tau_{2})\,\hat{a}^{\dagger}(\tau_{3})\,\hat{a}(\tau_{4})\right\rangle _{0},\notag\\
 \left\langle \hat{a}(\tau_{4})\,\hat{a}^{\dagger}(\tau_{1})\,\hat{a}^{\dagger}(\tau_{3})\,\hat{a}(\tau_{2})\right\rangle _{0}\,,&\left\langle \hat{a}(\tau_{4})\,\hat{a}(\tau_{2})\,\hat{a}^{\dagger}(\tau_{1})\,\hat{a}^{\dagger}(\tau_{3})\right\rangle _{0},\notag\\
 \left\langle \hat{a}(\tau_{4})\,\hat{a}^{\dagger}(\tau_{1})\,\hat{a}(\tau_{2})\,\hat{a}^{\dagger}(\tau_{3})\right\rangle _{0}\,,&\left\langle \hat{a}^{\dagger}(\tau_{1})\,\hat{a}(\tau_{4})\,\hat{a}(\tau_{2})\,\hat{a}^{\dagger}(\tau_{3})\right\rangle _{0},
\end{align}
With the help of the polariton mapping introduced in Ref.~\cite{key-2}, one can calculate these averages straightforwardly. Hence, we find for example the following expression for the expectation value
\begin{align}
 &\left\langle \hat{a}^{\dagger}(\tau_{1})\,\hat{a}(\tau_{2})\,\hat{a}^{\dagger}(\tau_{3})\,\hat{a}(\tau_{4})\right\rangle _{0}=\frac{1}{\mathcal{Z}_{0}}\sum_{n=1}^{\infty}\sum_{\alpha,\nu,\atop\rho,\pi=\pm}e^{-\beta\,E_{n\alpha}}\notag\\
 &\hspace{1em}\times e^{\left(E_{n\alpha}-E_{\left(n-1\right)\pi}\right)\tau_{1}}\,e^{\left(E_{\left(n-1\right)\pi}-E_{n\rho}\right)\tau_{2}}\;t_{n\alpha\pi}\; t_{n\rho\pi}\notag\\
 &\hspace{1em}\times e^{\left(E_{n\rho}-E_{\left(n-1\right)\nu}\right)\tau_{3}}e^{\left(E_{\left(n-1\right)\nu}-E_{n\alpha}\right)\tau_{4}}\;t_{n\rho\nu}\; t_{n\alpha\nu}\label{E:4OrderAverage_2_N1}\,.
\end{align}
Subsequently, performing a Matsubara transformation according to \eqref{E:DefinitionMatsubaraTrans} yields a formal integral of the form
\begin{align}
 I=\gamma \int _0^{\beta }dt e^{a t}\int _0^tdt_1e^{b t_1}\int _0^{t_1}dt_2e^{c t_2}\int _0^{t_2}dt_3e^{d t_3}
\end{align}
with the solution
\begin{align}\label{E:FormalIntegralSolution}
 I=&\gamma \Bigl[\frac{e^{(a+b+c+d) \beta }-1}{(a+b+c+d)(b+c+d)(c+d)d}\notag\\
 &-\frac{e^{(a+b+c) \beta }-1}{(a+b+c)(b+c)c d}+\frac{-1+e^{(a+b) \beta }}{b (a+b) c (c+d)}\notag\\
 &-\frac{-1+e^{a \beta }}{a b (b+c) (b+c+d)}\Bigr].
\end{align}
The variables $a,b,c,d$ correspond to differences of energy eigenvalues. Due to energy conservation these variables have to fulfill the condition $a+b+c+d=0$ and thus the above solution \eqref{E:FormalIntegralSolution} has a pole in the first term. Therefore, we have to take the limit
\begin{align}\label{E:ABCDLimit}
 \lim_{a+b+c+d\rightarrow 0}&\;\frac{e^{(a+b+c+d) \beta }-1}{(a+b+c+d)(b+c+d)(c+d)d}\notag\\
 &=\frac{\beta }{(b+c+d)(c+d)d}.
\end{align}
Since this pole arises for all expectation values \eqref{E:AllSix4PointCorrs} we always have to consider this particular limit. Taking this result into account, the explicit expression for the expectation value \eqref{E:4OrderAverage_2_N1} in Matsubara space is given by
\begin{widetext}
\begin{align}\label{E:A2explicit}
 &I_{\hat{a}^{\dagger}\hat{a}\hat{a}^{\dagger}\hat{a}}=\frac{1}{\mathcal{Z}_{0}\,\beta^2}\sum_{n=1}^{\infty}\sum_{\alpha,\nu,\atop\rho,\pi=\pm}e^{-\beta  E_{n,\alpha }} \Bigg\lbrace-\frac{\frac{-1+e^{\beta  \left(-\omega _{\text{m1}}-E_{-1+n,\lambda }+E_{n,\alpha }\right)}}{\left(\omega _{\text{m1}}+E_{-1+n,\lambda }-E_{n,\alpha }\right) \left(-\omega _{\text{m1}}-E_{-1+n,\lambda }+E_{n,\alpha }\right) \left(\omega _{\text{m2}}+E_{-1+n,\lambda }-E_{n,\rho }\right)}}{\omega _{\text{m2}}-\omega _{\text{m3}}+E_{-1+n,\lambda }-E_{-1+n,\nu }}\notag\\
 &+\frac{\frac{-1+e^{\beta  \left(-\omega _{\text{m4}}-E_{-1+n,\nu }+E_{n,\alpha }\right)}}{\left(\omega _{\text{m4}}+E_{-1+n,\nu }-E_{n,\alpha }\right) \left(-\omega _{\text{m4}}-E_{-1+n,\nu }+E_{n,\alpha }\right) \left(-\omega _{\text{m3}}-E_{-1+n,\nu }+E_{n,\rho }\right)}}{\omega _{\text{m2}}-\omega _{\text{m3}}+E_{-1+n,\lambda }-E_{-1+n,\nu }}+
 \frac{\frac{\beta}{\left(\omega _{\text{m1}}+E_{-1+n,\lambda }-E_{n,\alpha }\right) \left(\omega _{\text{m4}}+E_{-1+n,\nu }-E_{n,\alpha }\right)}}{-\omega _{\text{m3}}+\omega _{\text{m4}}-E_{n,\alpha }+E_{n,\rho }}\notag\\
 &+ \frac{\frac{-1+e^{\beta  \left(-\omega _{\text{m1}}+\omega _{\text{m2}}+E_{n,\alpha }-E_{n,\rho }\right)}}{\left(\omega _{\text{m2}}+E_{-1+n,\lambda }-E_{n,\rho }\right) \left(-\omega _{\text{m1}}+\omega _{\text{m2}}+E_{n,\alpha }-E_{n,\rho }\right) \left(-\omega _{\text{m3}}-E_{-1+n,\nu }+E_{n,\rho }\right)}}{-\omega _{\text{m3}}+\omega _{\text{m4}}-E_{n,\alpha }+E_{n,\rho }}\Bigg\rbrace \;t_{n,\alpha ,\lambda }\,t_{n,\alpha ,\nu } \,t_{n,\rho ,\lambda }\,t_{n,\rho ,\nu }\,.
\end{align}
\end{widetext}
However, this expression still possesses some poles for special choices of Matsubara frequencies. Fortunately, all these poles can be eliminated by investigating the corresponding limits analogous to equation \eqref{E:ABCDLimit}. Further care has to be taken considering the occurrence of the ground-state energy due to its uniqueness. Similar expressions can be calculated for the other expectation values \eqref{E:AllSix4PointCorrs}.

\end{document}